\documentclass[epsfig,12pt]{article}
\usepackage{epsfig,amssymb,euscript,bbold}
\usepackage{amsmath}
\newcommand{\be}{\begin{equation}}
\newcommand{\ee}{\end{equation}}
\newcommand{\bea}{\begin{eqnarray}}
\newcommand{\eea}{\end{eqnarray}}
\newcommand{\ba}{\begin{array}}

\newcommand{\ea}{\end{array}}

\def\bbox{{\,\lower0.9pt\vbox{\hrule \hbox{\vrule height 0.2 cm
\hskip 0.2 cm \vrule height 0.2 cm}\hrule}\,}}
\newcommand{\dsl}{\pa \kern-0.5em /}

\def\ds{\raise.15ex\hbox{/}\kern-.57em\partial}
\def\Ds{\,\raise.15ex\hbox{/}\mkern-13.5mu D}

\begin{document}

%%%%%%%%%%%%%%%%%%%%%%%%%%%%%%%%%%%%%%%%%%%%%%%%%%%%%%%%%%%%%%%%%%%%%%%%%%
\begin{titlepage}

\vfill

\begin{flushright}
hep-th/0411145\\
\end{flushright}

\vfill

\begin{center}

\baselineskip=16pt

{\Large\bf Half-BPS Giants, Free Fermions and Microstates of
Superstars}

\vskip 1.8cm

Nemani V. Suryanarayana

\vskip 1cm

{\small{\it Perimeter Institute for Theoretical
Physics\\ Waterloo, ON, N2L 2Y5, Canada\\ E-mail:
{\rm vnemani@perimeterinstitute.ca}\\}}

\end{center}

\vfill

\begin{center}
\textbf{Abstract}
\end{center}

We consider 1/2-BPS states in AdS/CFT. Using the matrix model
description of chiral primaries explicit mappings among configurations
of fermions, giant gravitons and the dual-giant gravitons are
obtained. These maps lead to a `duality' between the giant and the
dual-giant configurations which is the reflection of particle-hole
duality of the fermion picture. These dualities give rise to some
interesting consequences which we study. We then calculate the
degeneracy of 1/2-BPS states both from the CFT and string theory
and show that they match. The asymptotic degeneracy grows
exponentially with the conformal dimension. We propose that the
five-dimensional single charge `superstar' geometry should carry this
density of states. An appropriate stretched horizon can be placed in
this geometry and the entropy predicted by the CFT and the string
theory microstate counting can be reproduced by the Bekenstein-Hawking
formula up to a numerical coefficient. Similar M-theory examples are
also considered.

\begin{quote}

\end{quote}

\vfill

%\vfill \vskip 5mm \hrule width 5.cm \vskip 5mm

\end{titlepage}

%%%%%%%%%%%%%%%%%%%%%%%%%%%%%%%%%%%%%%%%%%%%%%%%%%%%%%%%%%%%%%%%%%%%%%%%%%

\section{Introduction}

Dualities in string theory have proved to be powerful tools in our
understanding of its physics. The most studied of these is the AdS/CFT
correspondence \cite{maldacena} (see \cite{agmoo} for a review and
references) which states that the ${\cal N}$ =4, d=4, SU(N) SYM CFT is
equivalent to the type IIB string theory on $AdS_5 \times S^5$. This
being a strong-weak duality to test it one usually relies on some
non-renormalisation theorems. In this context the 1/2-BPS operators of
the CFT, corresponding to the 1/2-BPS states on $S^3 \times R$ via the
state-operator correspondence, played a very important role as their
conformal dimensions are protected from quantum corrections. Under the
AdS/CFT correspondence these states are dual to 1/2-BPS states in the
type IIB string theory on $AdS_5 \times S^5$. However it is well known
that the chiral primary operators have many possible dual descriptions
on the string theory side. For small values of R-charge they are dual
to multiparticle supergravity/closed string states. As the R-charge
increases to $J \sim N$ the point like states are no longer good
descriptions and they are better described by large D3-branes with the
same R-charges. These D3-branes come in two classes: the giant
gravitons \cite{mst} and the dual-giant gravitons \cite{gmt,hhi}. The
giant gravitons are central to understanding the `stringy exclusion
principle' from the AdS point of view \cite{mst}. For $J \sim N^2$ new
geometries arise \cite{llm}.

Recently it has emerged that the correlators of chiral primaries in
the CFT and the physics of the corresponding dual objects in the bulk
are captured by a Hermitian matrix model with a harmonic oscillator
potential \cite{cjr, berenstein, llm}. In terms of the solution of
this matrix model each chiral primary operator corresponds to a
quantum mechanical configuration of $N$ fermions $\lambda_i$ in the
single particle spectrum of a harmonic oscillator.

In this paper we use the matrix model description to set up a
one-to-one correspondence between configurations of giant and
dual-giant gravitons. The duality follows from particle-hole duality
of the fermion picture. We also explain how stringy exclusion
principle manifests itself in terms of the dual-giants. Just as there
is an upper bound on the angular momentum of a single giant, it turns
out that there is an upper bound, namely $N$, on the number of
dual-giants. This result agrees nicely with the fermion picture and
the duality between giant and dual-giant configurations. We study some
of the interesting consequences of this novel picture.

Further we find the partition function and the asymptotic density of
the 1/2-BPS states from both the CFT and the string theory. On the CFT
side this amounts to counting the fixed energy configurations of $N$
fermions in the harmonic oscillator spectrum. On the string theory
side we count the giant graviton configurations which agrees with the
CFT result. We show that this density of states increases
exponentially with the total energy in the large-$N$ limit.

An exponential growth in the number of 1/2-BPS states with energy
prompts one to ask if there is a 1/2-BPS `black hole' in $AdS_5$ which
carries this density of states. However there are no 1/2-BPS
black holes known in $AdS_5$ with finite horizon area. We propose that
this degeneracy of states should be carried by the single charge
`superstar' geometry of \cite{mt}. This geometry arises as the
extremal limit of the non-supersymmetric single charge black hole
\cite{bcs1,bcs2} in $AdS_5$. When lifted to 10-dimensional type IIB
supergravity this solution preserves 16 supercharges and admits an
interpretation as the backreacted geometry of a specific configuration
of giant gravitons on $S^5$ (and thus denoted the `superstar' in
\cite{mt}). Since it preserves 16 supersymmetries in 10 dimensions one
expects that its microstates should be dual to the 1/2-BPS operators
of the ${\cal N}=4$, d=4 SU(N) SYM on the boundary.

The single charge superstar geometry has a null
singularity. Classically such singularities can be thought of as
black holes with zero size horizons. Well studied examples of such
geometries with flat asymptotes arise as the extremal limits of two
charge black holes in 4 and 5 dimensions which are related to the
physics of D1-D5 system. Even though classically the area of the
horizon is zero, one expects big quantum corrections to the
Bekenstein-Hawking formula, {\it e.\,g}, \cite{sen, wald, cdm1,cdm2,
dabholkar}. Alternatively such singularities are expected to get
corrected into black holes of finite size horizon once the quantum
corrections are included \cite{dkm}.

Motivated by the arguments advocated by Mathur {\it et al} (see
\cite{lm} for instance) in recent times in the context of D1-D5
systems, we place a stretched horizon in the single charge superstar
geometry in 5 dimensions and show that the Bekenstein-Hawking formula
for the entropy reproduces the predicted answer up to a non-zero
number.

A similar analysis is carried out for the null singularities that
arise in 4 and 7 dimensional gauged supergravities as well. These
solutions when lifted to 11-dimensional supergravity are
asymptotically $AdS_4 \times S^7$ and $AdS_7 \times S^4$ respectively
and preserve 16 supercharges. These again admit interpretation as the
backreactions of giants in M-theory \cite{bn,lmp} now made of M2 and
M5 branes. We exhibit the matching of gauge/M-theory prediction for
the entropy with that of the stretched horizon picture.

The rest of the paper is organized as follows. In the Section 2 we
review some aspects of 1/2-BPS states in AdS/CFT and the matrix model
description of these states. We argue that $N$ is the upper bound on
the number of dual-giants. In Section 3 we describe the above
mentioned `duality' between configurations of giants and dual-giants
using the matrix model description. In Section 4 we find the partition
function of the 1/2-BPS states using the fermion picture in the CFT
and counting the giant graviton configurations on the string theory
side. We use it to calculate the asymptotic density of states in the
large-$N$ limit and show that it grows exponentially with the
conformal dimension (equivalently the R-charge). Section 5 contains a
review of the geometry of the single charged `superstar' and some of
its features relevant for us. In Section 6 we place a stretched
horizon and recover the predicted entropy up to a number. Section 7
contains similar results for the M-theory superstars. In Section 8 we
discuss some more consequences of the duality between the giant and
the dual-giant configurations. We end with some concluding remarks in
Section 9.\footnote{Some of the results of Sections 2 and 3 were
previously obtained in \cite{bs} using different methods. We thank
Iosif Bena for pointing out \cite{bs}.}

%\newpage

\section{Chiral primaries and giant gravitons}

Let us start by briefly reviewing the relevant information of the
${\cal N} = 4$, d=4, SU(N) SYM theory which is dual to type IIB string
theory on $AdS_5 \times S^5$ background \cite{maldacena} (see, {\it
e.\,g.,} \cite{agmoo} for a review). This theory has a large number of
half-BPS operators, namely, the chiral primaries. These operators
belong to $(0,l,0)$ representation of the R-symmetry group $SU(4)\sim
SO(6)$. In terms of ${\cal N}=1$ notation, the ${\cal N} =4$ theory has
three chiral multiplets and a vector multiplet. A generic chiral
primary can be written as:
\begin{equation}
\label{mtrcp}
(tr(\Phi^{l_1}))^{k_1} (tr(\Phi^{l_2}))^{k_2} \cdots
(tr(\Phi^{l_m}))^{k_m}
\end{equation}
where $\Phi$ is the complex scalar of one of the three chiral
multiplets. The conformal dimension $\Delta$ of these operators equals
their R-charge $J$
\begin{equation}
\Delta = J
\end{equation}
Supersymmetry protects the conformal dimensions of chiral primaries
from receiving quantum corrections. The operator in (\ref{mtrcp}) has
\begin{equation}
\label{deltaofcp}
\Delta = \sum_{i=1}^m l_i k_i.
\end{equation}
For finite $N$ the operators of type (\ref{mtrcp}) are independent
only if $l_m \le N$. The above basis in Eq.\,(\ref{mtrcp}) is
approximately orthonormal in the large-$N$ limit \cite{berenstein}. A
different orthonormal basis called the `Schur polynomial basis' was
introduced in \cite{cjr}. A Schur Polynomial $\chi_R (\Phi)$ is the
character of the unitary group in a given irreducible representation
$R$. Since the irreducible representations of the unitary group can be
represented by Young tableau one has a Schur polynomial for each Young
tableau. The total number of boxes in a Young diagram gives the total
R-charge $J$ of the corresponding chiral primary.

The correlation functions of chiral primaries in this basis have been
calculated to all order in 1/$N$ and at the tree level in $\lambda$,
the 't Hooft coupling in \cite{cjr}(see also \cite{kg}). There, a
Hermitian matrix model with a harmonic oscillator potential was
proposed to capture the correlation functions of chiral
primaries. This matrix model can be obtained as a truncation of the
d=4 SYM on $S^3 \times R$ down to the zero modes over $S^3$ in the
weak coupling and keeping just one complex scalar $\Phi$
\cite{berenstein}. This model was further studied in \cite{berenstein,
cs, llm}. Some relevant features of this are briefly reviewed below.

\subsection{The matrix model description of chiral primaries}

As it has been explained in \cite{cjr,berenstein} (see also \cite{im})
the matrix model can be reduced to the quantum mechanics of $N$
fermionic eigenvalues $\lambda_i$ of the matrix in a harmonic
oscillator potential. The Hamiltonian of this system is
\begin{equation}
\label{fermiham}
H = \sum_{i=1}^N (\lambda_i^\dagger \lambda_i + \frac{1}{2}).
\end{equation}
Each stationary state of this quantum mechanics is given by a
configuration of $N$-fermions in the harmonic oscillator energy
spectrum $E_j = j + 1/2$, $j = 0,1,\cdots$. Since there are $N$
fermions the vacuum energy is
\begin{equation}
E_0 = \frac{1}{2} + \frac{3}{2} + \cdots + \frac{2N-1}{2} =
\frac{N^2}{2}. 
\end{equation}
The Hilbert space is spanned by $N$-particle states labeled by $N$
single particle levels $f_k$ or an $N$-vector ${\vec f} = (f_1, f_2,
\cdots , f_N)$ where $0 \le f_1 < f_2 < \cdots < f_N$. The eigenvalue
of the $N$-particle Hamiltonian (\ref{fermiham}) on this state is $E =
\frac{N}{2} + \sum_{k=1}^N f_k$. We choose to measure the energies of
the states to be the difference of the full energy and the ground
state energy: $\Delta = E - E_0$.

Note that each excitation of the fermion system can be mapped uniquely
to a U($N$) Young tableau and thus a representation $R$ of
U($N$). This is done by mapping the particle excitations to successive
rows of the tableau. In our case since there are $N$ fermions, the
Young tableau contains a maximum of $N$ rows. The number of boxes in
the k$^{th}$ row is $f_k - (k-1)$. Therefore each chiral primary
written in the Schur polynomial basis corresponds to a unique
configuration of the fermion system \cite{cjr, berenstein}. We will
make use of this fermion picture below.

Let us next turn to describing the string theory duals of chiral
primaries.

\subsection{Giants and dual-giants}

As mentioned in the introduction the chiral primary operators have
many possible dual descriptions on the string theory side. For small
values of R-charge they are dual to supergravity modes. However as the
R-charge increases $J \sim N$ the duals are better described by
D3-branes with the same R-charges. Further, these D3-branes come in
two classes: the giant gravitons \cite{mst} and the dual-giant
gravitons \cite{gmt,hhi}. We will be interested in counting the duals
of chiral primaries later on. So to avoid over-counting we have to
restrict ourselves to either counting the supergravity KK modes or the
giant D3-branes. Since we are interested in states with very large
R-charge we choose the giant D3-brane basis. It turns out that we
should not count giant and dual-giant configurations separately as
there is a `duality' relating both which we will explain below. We
will also argue that there is an upper bound, given by $N$, to the
total number of dual-giants.

Before doing this let us review some relevant aspects of the giant and
the dual-giant gravitons in type IIB string theory on $AdS_5 \times
S^5$. Giant gravitons \cite{mst} are D3-branes wrapping an $S^3$
inside the $S^5$ and rotating along one of the transverse directions
within the $S^5$. Since they do not wrap any homological cycle they do
not carry any net D3-brane charge, but they do have a D3 dipole
moment. They preserve 16 of the 32 supersymmetries\footnote{There are
also giant gravitons that carry more than one R-charge which are 1/4
or 1/8 supersymmetric \cite{mikhailov}. We will not consider these
configurations here.} of $AdS_5 \times S^5$.

To be more specific let us work with the following coordinate system
for $AdS_5 \times S^5$ in global coordinates.
\begin{equation}
\label{globalcoords}
ds^2_{AdS_5} = - \left( 1+\frac{r^2}{L^2} \right) dt^2 + \frac{dr^2}{
1+\frac{r^2}{L^2}} + r^2 d\Omega_3^2,
\end{equation}
\begin{equation}
ds^2_{S^5} = L^2 \left( d\alpha^2 + \sin^2\alpha \, d\beta^2 + \cos^2
\alpha \, d\xi_1^2 + \sin^2\alpha \, \left[ \cos^2 \beta \, d\xi_2^2 +
\sin^2\beta \, d\xi_3^2 \right] \right).
\end{equation}
Here the ranges of the coordinates are: $0 \le r < \infty$, $ 0 \le
\alpha, \beta \le \pi/2$, and $0 \le \xi_i \le 2\pi$ for
$i=1,2,3$. Further assume that the D3-brane wraps the $S^3 \subset
S^5$
\begin{equation}
ds^2_{S^3} = L^2 \sin^2 \alpha \left[ d\beta^2 + \cos^2 \beta \,
d\xi_2^2 + \sin^2\beta \, d\xi_3^2  \right]
\end{equation}
and rotates along the $\xi_1$ direction. The angular momentum of a
single such D3-brane is given by
\begin{equation}
P_{\xi_1} = N \sin^2 \alpha.
\end{equation}
Thus we have $P_{\xi_1} \le N$ which realises the stringy exclusion
principle. In a quantum theory one expects that $P_{\xi_1}$ is
quantised. Hence the allowed values of $\alpha$ would be discrete such
that $P_{\xi_1}$ takes values $0,1,2, \cdots, N$. A general
configuration of giants is then given by an $N$-vector ${\vec b}_1 =
(r_1, r_2, \cdots, r_N)$ where the integers $r_k \in [0, \infty)$
denote the number of giant gravitons with angular momentum $P_{\xi_1}
= k$. The total energy (and therefore the angular momentum) of this
configuration is $\sum_{k=1}^N k \, r_k$. An individual giant graviton
with R-charge $m$ is dual to the subdeterminant operator \cite{bbns}:
\begin{equation}
\label{subdet}
\frac{1}{m!} \epsilon_{i_1\cdots i_m i_{m+1} \cdots i_N} \epsilon^{j_1
\cdots j_m i_{m+1} \cdots i_m} \Phi^{i_1}_{j_1} \cdots \Phi^{i_m}_{j_m}.
\end{equation}
The fact that these operators do not exist for $m > N$ is again the
stringy exclusion principle. This operator (\ref{subdet}) is a special
element in the Schur polynomial basis and corresponds to a Young
tableau with a single column with $L$ boxes in it.
 
On the other hand the dual-giant gravitons \cite{gmt, hhi} are
D3-branes wrapping $S^3 \subset AdS_5$ and rotating along a maximal
circle of $S^5$. They again preserve 16 supercharges of the background
and carry D3-brane dipole moment. For a given angular momentum
$P_{\xi_1}$ the radius $r$ at which the D3-brane stabilises is given
by
\begin{equation}
r = L \, \sqrt{\frac{P_{\xi_1}}{N}}
\end{equation}
where $L$ is the radius of $AdS_5$. Since the radial coordinate $r$
ranges from 0 to $\infty$ the dual-giants can have arbitrary (integer
valued) angular momenta. 

This raises the question of how the stringy exclusion principle
manifests itself for the dual-giants. To answer this let us note a
subtle effect which restricts the total number of dual-giants that one
can place in $AdS_5 \times S^5$ (see also \cite{bs}). Since a
dual-giant occupies three of the four spacelike coordinates in
$AdS_5$, it acts like a domain-wall and the flux of $F^{(5)}$ measured
on either side of this domain-wall differs by one unit with the lesser
value on the inside of $S^3$ that the dual-giant wraps \cite{hhi}. So
if we have $m$ dual-giants in $AdS_5$ the $F^{(5)}$ flux measured
inside the inner most dual-giant will be $N-m$ units. For $m = N$ the
five form flux inside the innermost dual-giant vanishes. Since it is
crucial to have non-zero flux to stabilise a dual-giant at a non-zero
radius and to produce a geometry where there are closed
orbits\footnote{I thank Rob Myers for suggesting this argument.}, it
follows that we can not have any more dual-giants in the system. This
is the manifestation of `stringy exclusion principle' for the
dual-giants.

Taking this into account a general configuration of dual-giants is
also given by an $N$-vector ${\vec b}_2$ = ($s_1$, $s_2$, $\cdots$,
$s_N)$. Here the integers $s_k$ are such that $0 \le s_N \le \cdots
\le s_1 < \infty$ and $s_k$ denotes the angular momentum of the
k$^{th}$ dual-giant away from the boundary of $AdS_5$. The total
energy of this configuration is given by $H_{\vec b_2} = \sum_{k=1}^N
s_k$.

Next we propose a one-to-one map among configurations of giants and
configurations of dual-giants using the fermion picture coming from
the CFT.

\section{Mapping fermions and branes}

To summarise, a configuration of the $N$ fermion system is specified
by an $N$-vector ${\vec f} = (f_1$, $f_2$, $\cdots$, $f_N)$ with $0\le
f_1 < f_2 \cdots f_N \le \infty$ and $f_i$ denoting the level number
of the i$^{th}$ fermion. The total energy of such a configuration is
\begin{equation}
H_{\vec f} = -\frac{N(N-1)}{2} + \sum_{k=1}^N f_k.
\end{equation}
A configuration of giant gravitons is specified by another $N$-vector
of integers ${\vec b}_1 = (r_1$, $r_2$, $\cdots$, $r_N)$ with $0\le
r_k \le \infty$. Each $r_k$ denotes the number of giant gravitons at
level-$k$ in the $N$-level system. The energy of this configuration is
\begin{equation}
H_{{\vec b}_1} = \sum_{k=1}^N k r_k.
\end{equation}
The dual-giant graviton system is described by yet another $N$-vector
of integers ${\vec b}_2 = (s_1, s_2, \cdots, s_N)$ with $s_1 \ge s_2
\ge \cdots \ge s_N \ge 0$. Each $s_k$ denotes the angular momentum of
the k$^{th}$ dual-giant D3-brane away from the boundary of AdS. The
energy of this system is
\begin{equation}
H_{{\vec b}_2} = \sum_{k=1}^N s_k.
\end{equation}
A fermion configuration can be specified either in terms holes or
particles. We first observe that each hole can be associated with a
giant graviton and each particle with a dual-giant graviton. For
instance a hole at the level-0 (the ground state of the single
particle harmonic oscillator Hilbert space) can be thought of as the
one in which all the $N$ particles are excited by one energy level
each. As explained in \cite{berenstein}, a hole at the level-0 is dual
to a single giant graviton with the maximum possible angular momentum
$N$. On the other hand in terms of dual-giants this excitation
corresponds to having $N$ dual-giants with one unit of angular
momentum each. Of course the description in terms of a single giant is
a better one as the probe brane approximation will be valid. Similarly
consider an excitation of the topmost fermion by $n$ levels. This, as
in \cite{berenstein} can be thought of as a dual-giant with $n$ units
of angular momentum. But this can also be associated with a
configuration of $n$ smallest size giant gravitons. Again the former
would describe the system better than the latter. Similarly one can
analyse a given configuration of fermions and associate a unique giant
or a dual-giant configuration with it. Doing this we find the
following maps between the fermion configurations and the giant and
the dual-giant configurations:
\begin{eqnarray}
\label{themaps}
r_N &\leftrightarrow& f_1, ~~ r_{N-i} \leftrightarrow f_{i+1} - f_i
- 1, ~~ i = 1, 2, \cdots, N-1 \cr
s_{N-i} &\leftrightarrow& f_{i+1} - i, ~~ i = 0,1,\cdots,N-1
\end{eqnarray}
Clearly under these identifications the Hamiltonians of the systems
match along with the restrictions on the vectors ${\vec f}$, ${\vec
b}_1$ and ${\vec b}_2$. Since there is a single fermion configuration
for either of the two bosonic (giant and dual-giant) configurations it
is natural to conjecture that the two bosonic systems should be `dual'
to each other with the following mapping:
\begin{equation}
\label{themapstwo}
s_i \leftrightarrow \sum_{k=i}^N r_k,  ~~ i = 1,2,\cdots,N 
\end{equation}
This `duality' map of (\ref{themapstwo}) between the configurations of
giants and dual-giants is a direct result of the duality between
descriptions of the fermion system in terms of particles or
holes.\footnote{The relation in Eq.\,(\ref{themapstwo}) for the case
of single non-vanishing $r_k$ was proposed earlier in \cite{bs}.}
Recall that each fermion configuration is uniquely determined by a
Young tableau. Using (\ref{themaps}) the number of boxes in each row
can be identified with the corresponding $s_k$. Therefore a given
Young tableau can be associated uniquely to a specific configuration
of dual-giants with the number of boxes in the k$^{th}$ row giving the
angular momentum of the corresponding dual-giant. The fact that there
are a maximum of $N$ rows reflects the fact that there is an upper
bound on the number of dual-giants. This is the particle description
of the fermion system.

On the other hand the same fermion configuration can be equivalently
described by holes. And the hole excitations can also be described by
the same Young tableau. This picture can be associated with a
configuration of giants with each column representing a giant graviton
and the number of boxes in that column being its angular
momentum. This type of a duality when applied to M-theory context
would mean that some configurations of giants made of M5 (M2) branes
in $AdS_4 \times S^7$ ($AdS_7 \times S^4$) are dual to the
corresponding configurations of dual-giants made of M2 (M5) branes. A
similar duality has already been proposed by \cite{msv} in the context
of M-theory in a pp-wave background.

Even though there are dualities between different descriptions of
chiral primaries on the string theory side, one has to keep in mind
that for most of the situations only one of the three candidates,
namely the point like KK modes, the giant gravitons or the dual-giant
gravitons, is a good description but not all. In some cases none of
them alone describes the true physics in which case one has to work
with the full supergravity solution \cite{llm}. We next turn to
enumerating the 1/2-BPS states and finding their asymptotic
degeneracies for fixed $\Delta$ ($J$).

\section{Asymptotic densities of 1/2-BPS states}

One can write down a generating function to summarise the number of
independent chiral primary operators for a given $\Delta$. One can
check that the function
\begin{equation}
\label{partfnone}
Z_N(q) = \prod_{n=1}^N (1-q^n)^{-1} = \sum_{\Delta = 0}^\infty
d_{\Delta} q^\Delta
\end{equation}
fits the bill where $q = e^{-\beta}$ and $d_\Delta$ gives the number
of independent chiral primaries with conformal dimension
$\Delta$. This partition function can be calculated in many
ways. Before analysing (\ref{partfnone}) further for the asymptotic
density of states let us derive this from the matrix model by counting
the fermion configurations and from string theory by counting the
configurations of giant gravitons.

\subsection{Partition function from the matrix model}

The partition function of the system of $N$ fermions in a harmonic
oscillator potential is \cite{im}:
\begin{equation}
\label{fdpartone}
Z_N(q) = q^{-\frac{N^2}{2}}{\rm Tr} \, q^H =
q^{-\frac{N^2}{2}}\sum_{f_1=0}^\infty 
\sum_{f_2 = f_1+1}^\infty 
\cdots \sum_{f_N = f_{N-1}+1}^\infty q^{\sum_{n=1}^N (f_n +
\frac{1}{2})}
\end{equation}
with $q = e^{-\beta}$. To perform the sums we make the following
change of variables: $r_N = f_1$, $r_{N-i} = f_{i+1} - f_i -1$ for
$i=1, 2, \cdots, N-1$. In terms of the new variables
Eq.\,(\ref{fdpartone}) becomes
\begin{equation}
Z_N(q) = \sum_{r_1 =0}^\infty \sum_{r_2 = 0}^\infty \cdots \sum_{r_N =
0}^\infty q^{\sum_{j=1}^N j\, r_j}
\end{equation}
which can be summed easily to get
\begin{equation}
\label{partfntwo}
Z_N (e^{-\beta}) = \sum_J d_\Delta e^{-\beta \Delta} = 
\prod_{n=1}^N \frac{1}{1-q^n}.
\end{equation}
Here $\Delta = J$ is again the total $U(1)$ R-charge and $d_{\Delta}$
is the degeneracy of states with fixed $\Delta$. So we have recovered
the partition function of chiral primaries (\ref{partfnone}) using the
matrix model.

\subsection{Partition function from giants}

As discussed earlier in Section (2.2) there are three different
known duals of 1/2-BPS states of the SYM. We choose to count the giant
and the dual-giant configurations. Invoking the `duality' of Section 3
between configurations of giants and dual-giants we can simply choose
to count giant graviton configurations. It is clear that counting the
dual-giants will give identical results.

It is known from \cite{djm} that there are no 1/2-BPS fluctuations of
the giant gravitons and therefore it is sufficient to treat these as
simple particles. Thus the problem of counting the giant graviton
configurations reduces to that of counting the configurations of
bosons in an equally spaced $N$-level system. As explained earlier a
general configuration of the giants is labeled by $N$ integers ${\vec
b}_1 = (r_1, r_2, \cdots, r_N)$. The total R-charge of such a
configuration is:
\begin{equation}
P_{\xi_1} | {\vec b}_1 \rangle = \left( \sum_{k=1}^N k r_k
\right) | {\vec b}_1 \rangle.
\end{equation}
Thus counting the number of giant graviton configurations with a fixed
total angular momentum again reduces to the problem of partitions of
an integer $Q = \sum_{k=1}^N k r_k$. The partition function of this
problem is
\begin{eqnarray}
Z_N(q) &=& \sum_{r_1, \cdots, r_N} q^{\sum_{k=1}^N
k r_k} \cr 
&=& \prod_{k=1}^N \left( \sum_{r_k=0}^\infty q^{k r_k} \right)
= \prod_{k=1}^N \frac{1}{1-q^k}
\end{eqnarray}
where $q = e^{-\beta}$. Of course this is precisely the same partition
function Eq.\,(\ref{partfntwo}) obtained from the CFT (matrix model)
side.

\subsection{The asymptotic degeneracy and the entropy}

To further analyse $Z_N(q)$, define the free energy
\begin{eqnarray}
\label{fenergy}
F(q,N) &=& \ln Z_N(q) = -\sum_{n=1}^N \ln (1-e^{-\beta \, n}) \cr
&=& -\sum_{n=1}^\infty \ln (1-e^{-\beta \, n}) + \sum_{n=N+1}^\infty
\ln (1-e^{-\beta \, n}) \cr
&=& \ln Z_\infty (q) + \sum_{p=1}^\infty \frac{1}{p} \,
\frac{e^{-\beta(N+1)p}}{1-e^{-\beta p}}. 
\end{eqnarray}
Using Eq.\,(\ref{fenergy}) we can write $Z_N(q)$ as
\begin{eqnarray}
\label{znapprox}
Z_N(e^{-\beta}) &=& \frac{e^{-\frac{\beta}{24}}}{\eta(e^{-\beta})}
~\exp\left[\sum_{p=1}^\infty \frac{1}{p} \,
\frac{e^{-\beta(N+1)p}}{1-e^{-\beta p}} \right] \cr
&=& \frac{e^{-\frac{\beta}{24}}}{\eta(e^{-\beta})} \left[ 1 + {\cal
O}(e^{-\beta \, (N+1)}) \right]
\end{eqnarray}
where $\eta(q)$ is the standard Dedikind's eta function:
\begin{equation}
\eta (q) = q^{\frac{1}{24}} \prod_{n=1}^\infty (1 - q^n).
\end{equation}
In the large-$N$ limit, we can neglect the ${\cal O}(e^{-\beta \,
(N+1)})$ corrections to $Z_N(q)$ and we will do so for the rest of
this paper.

From Eq.\,(\ref{partfnone}) $d_\Delta$ is given by:
\begin{equation}
\label{dddef}
d_\Delta = \frac{1}{2\pi i} \int_C d\beta \,e^{\beta \Delta}\,
Z_N (e^{-\beta}) 
\end{equation}
We are interested in extracting $d_{\Delta}$ for large-$\Delta$ and
$N$. Substituting the leading value of $Z_N(q)$ from
Eq.\,(\ref{znapprox}) into Eq.\,(\ref{dddef}) we have
\begin{equation}
d_\Delta = \frac{1}{2\pi i} \int_C d\beta  \frac{e^{\beta (\Delta -
\frac{1}{24})}} {\eta(e^{-\beta})}.  
\end{equation}
To find the asymptotic density $d_\Delta$ for large $\Delta$ we need
to take the `high temperature' limit $\beta \rightarrow 0$. It is
convenient to make a modular transformation of
$\eta(q)$. Recall:
\begin{equation}
\eta(e^{-\beta}) = \sqrt{\frac{2 \pi}{\beta}} \,
\eta(e^{-4\pi^2/\beta}). 
\end{equation}
Using this, $d_\Delta$ can be rewritten as
\begin{equation}
d_\Delta = \frac{1}{2\pi i} \int_C d\beta  \, e^{\beta (\Delta -
\frac{1}{24})} \left( \frac{\beta}{2\pi} \right)^{1/2} 
\frac{1}{\eta(e^{-4\pi^2/\beta})}. 
\end{equation}
Now use $\eta(q) \sim q^{\frac{1}{24}}$ as $q
\rightarrow 0$ to obtain:
\begin{equation}
\label{approxddelta}
d_\Delta = \frac{1}{2\pi i} \int_C d\beta  \, e^{\beta (\Delta -
\frac{1}{24}) +\frac{\pi^2}{6\beta} } \left( \frac{\beta}{2\pi}
\right)^{1/2} .
\end{equation}
In the saddle point approximation (\ref{approxddelta}) evaluates to
\begin{equation}
\label{density}
d_\Delta \approx \frac{1}{4 \sqrt{3} ~ \Delta} \, e^{\pi
\sqrt{\frac{2\Delta}{3}}}. 
\end{equation}
Thus the density of the 1/2-BPS states grows exponentially. From
Eq.\,(\ref{density}) the Boltzman's entropy formula $S = \ln d_\Delta$
gives
\begin{equation}
\label{sprediction}
S = \left( \frac{2\pi^2}{3} \right)^{1/2} ~ \sqrt{\Delta} ~ +
\cdots 
\end{equation}
where $\cdots$ are the corrections negligible in the large-$N$ and
large-$\Delta$ limit (with $N>>\sqrt{\Delta}$) which we drop
henceforth.

Thus we see that if there is a physical system with the 1/2-BPS states
as its microstates then it is expected to have a macroscopic entropy
for large $\Delta$. So it is natural to ask whether there is a
candidate in the bulk $AdS_5$ which carries this entropy. The dual
should be asymptotically $AdS_5$, should preserve 16 supercharges when
lifted to a 10-dimensional solution and carry just one U(1)
R-charge. There are no BPS black holes with these properties and with
finite size horizons in $AdS_5$. Instead in what follows we propose
that the single charge `superstar' of \cite{mt} should have this
entropy.

\section{The R-charge black holes in $AdS_5$ and the superstar}

Let us review the relevant `superstar' solution of ${\cal N} = 2$, d =
5 gauged supergravity and its 10-dimensional lift. The bosonic field
content of the 5-dimensional gauged supergravity is the metric, two
scalars parametrised by $X_i$, $i$ = 1, 2, 3 satisfying $X_1 X_2 X_3 =
1$ and three abelian gauge fields $A_i$. The theory admits
non-extremal charged black holes \cite{bcs1, bcs2}:
\begin{eqnarray}
\label{fivedbhone}
ds_5^2 &=& -(H_1 H_2 H_3)^{-2/3} f dt^2 + (H_1H_2H_3)^{1/3} (f^{-1}
dr^2 + r^2 d\Omega_3^2), \cr
X_i &=& H_i^{-1} (H_1H_2H_3)^{1/3}, ~~ A_i = (H_i^{-1} -1) dt, 
\end{eqnarray}
where
\begin{equation}
f = 1 - \frac{\mu}{r^2} + \frac{r^2}{L^2} H_1H_2H_3, ~~ H_i = 1 +
\frac{q_i}{r^2}. 
\end{equation}
Here $\mu$ is the non-extremality parameter, $L$ is the radius of the
asymptotic $AdS_5$ and $q_i$ determine the independent $U(1)^3$
charges. As noted in \cite{mt}, the area of the horizon shrinks as
$\mu$ decreases. There is a qualitative difference between the case
when only one charge is non-vanishing , {\it e.\,g.}, $q_1 \ne 0, \,
q_2 = q_3 = 0$, and the case when more than one charges are
non-vanishing. If only $q_1$ is non-zero then the horizon shrinks to
zero size precisely when $\mu \rightarrow 0$ and the solution becomes
BPS with a null singularity. For the other cases the horizon shrinks
to zero size for some nonzero $\mu = \mu_{crit}$ leaving behind a
naked timelike singularity as $\mu \rightarrow 0$.

Since we are interested only in the extremal solution and with just
one U(1) R-charge we will set $q_1 = Q$, $\mu = q_2 = q_3 =0$ where $Q
\ne 0$. This single charge BPS solution lifts to the following
solution of 10-dimensional type IIB supergravity \cite{cdhjl3mpst,
mt}.
\begin{eqnarray}
\label{tendmetric}
ds_{10}^2 &=& \sqrt{D} \Big[ - H^{-1} \left(1 + \frac{r^2}{L^2}H
\right) dt^2 + \frac{dr^2}{1 + \frac{r^2}{L^2}H} + r^2 d\Omega_3^2
\Big], \cr 
&+& \frac{1}{\sqrt{D}} \Big[ H (L^2 d\mu_1^2 + \mu_1^2 [L
d\xi_1 + (H^{-1} -1) dt]^2 ) + \sum_{i=2}^3 L^2 (d\mu_i^2 +
\mu_i^2 d\xi_i^2) \Big]
\end{eqnarray}
where $D = \mu_1^2 + \frac{1}{H} (\mu_2^2 + \mu_3^2)$, $\mu_1 = \cos
\alpha$, $\mu_2 = \sin\alpha \cos\beta$, $\mu_3 = \sin\alpha \sin
\beta$ and $H = 1+Q/r^2$. We will be working with $Q/L^2 <<1$. This
metric is supported by the 5-form field strength $F^{(5)} = dB^{(4)} +
*dB^{(4)}$ where
\begin{equation}
B^{(4)} = -\frac{r^4}{L} D \, dt \wedge d^3\Omega - L Q \mu_1^2 (L
d\xi_1 - dt) \wedge d^3\Omega.
\end{equation}
This geometry admits 16 supersymmetries. By looking at the dipole
moment of $F^{(5)}_{\alpha \beta \theta \phi \psi}$ where
$\theta,\phi,\psi$ are the coordinates on the $S^3 \subset AdS_5$ the
authors of \cite{mt} concluded that the 10-dimensional solution
(\ref{tendmetric}) can be interpreted as the condensate of giant
gravitons with angular momentum along $\xi_1$ and with a density of
branes along $\alpha$ given by
\begin{equation}
\label{giantdensity}
dn = N \frac{Q}{L^2} \sin 2\alpha \,
d\alpha.
\end{equation}
The total number of giant gravitons is 
\begin{equation}
n_{tot} = N \frac{Q}{L^2}
\end{equation} 
and the total angular momentum of the system is
\begin{equation}
\label{sugraangmon}
P_{\xi_1} =
\frac{N^2}{2} \frac{Q}{L^2}.
\end{equation}
In terms of the giant graviton configurations counted in Section 3,
the density of giants in Eq.\,(\ref{giantdensity}) is given by $r_k =
Q/L^2$ for all $k$ in the continuum limit.\footnote{This
interpretation can be translated into the fermion picture using
(\ref{themaps}) and recover the phase space diagram proposed in
\cite{llm,gh} corresponding to Eq.\,(\ref{tendmetric}) in the
continuum limit.}

This geometry was named the `superstar' in \cite{mt}. This solution
satisfies all our requirements except that it has an uncloaked null
singularity. Nevertheless one may think of the 5-dimensional geometry
as a black hole with a zero size horizon. On general grounds one
expects that the Bekenstein-Hawking entropy formula for this geometry
to get corrected in the quantum theory. As in the recent developments
in the asymptotically flat 2-charge extremal black holes
\cite{dabholkar, dkm} it is conceivable that one recovers the finite
entropy for the superstar geometry too after quantum
corrections. However this analysis is outside the scope of this
paper. Instead what we do here is to follow the reasoning of
\cite{sen,lm,mathur} and place a stretched horizon in our geometry.

\section{Stretched horizon and the entropy of superstar}

Recall that the total R-charge of the single charge superstar is
\begin{equation}
Q_{total} = \frac{N^2 Q}{2L^2}
\end{equation}
Substituting this into Eq.\,(\ref{sprediction}) we get the entropy
prediction for the 1/2-BPS black hole coming from both the CFT and
string theory:
\begin{equation}
\label{theprediction}
S_{BH} = \left( \frac{\pi}{\sqrt{3}} \right) N \sqrt Q_0
\end{equation}
where we have defined $Q_0 = Q/L^2$. For large $N$ and for any finite
value of $Q_0$ this entropy is macroscopic. Next we would like to
reproduce this answer using the mechanism of stretched horizon for the
single charge superstar geometry.

A stretched horizon in the context of heterotic string compactified
down to four dimensions has been defined in \cite{sen} as the place
where the string world-sheet becomes strongly coupled. A different
point of view has recently been advocated by Mathur and collaborators
(see \cite{lm,mathur} for instance). Here the basic idea is that the
brane system that makes up the black hole has a nonzero size. Further,
there are some excitations of the brane system which makes up the
black hole that can spread over long distances compared to string
length. One has to place a stretched horizon so that these
fluctuations are inside it. Typically this length scale is set by the
smallest possible excitation of the system.

In our case each microstate of the superstar geometry can be thought
of as a giant graviton configuration with fixed total R-charge. This
system is also expected to have a finite size. The source of this
finite size can be traced to the configuration of fermions in the
phase space representation of \cite{berenstein, llm}. There, one may
think of the system as an incompressible fluid. Any excited state
requires creating holes in the fermion droplet and therefore the
droplet spreads over a larger area. This spreading directly translates
into a spread in the radial direction in the AdS space using the
results of \cite{llm}. To estimate the extent of a given microstate
let us note that the lightest excitation of the giant graviton system
is to add a single unit of angular momentum. This can be done by
adding a single giant graviton to any of the microstate configurations
of giants at level-1 (or shift a giant graviton from level $k$ to
$k+1$). We have to convert this energy into a length scale. Roughly
speaking if we use this energy to create a dual-giant graviton then
the length scale that sets the size of this dual-giant, in this case
given by $L/\sqrt{N}$, should also set the size of the stretched
horizon.

From the point of view of the fermion system the smallest excitation
is to excite the topmost fermion by one energy level.\footnote{This
excitation can be thought of as creating a smallest size giant
graviton or equivalently a smallest size dual-giant with a unit of
R-charge as in Section 3.}  It is easy to see from the analysis of
\cite{llm} that to accomplish this operation it costs energy of order
$N$ and the disturbance of this excitation is $\delta R \sim L^2/N$
where $R = \mu_1 L^2(1+r^2/L^2)^{1/2}$ \cite{llm} with $R$ being the
radial coordinate in the phase space and $\mu_1 = \cos\alpha$. So we
see that supergravity solutions of two configurations of giants which
differ by this excitation roughly differ from each other within the
range of $ 0 \le r \le c_0 L/\sqrt{N}$ where $c_0$ is a pure number.

Following the analogy of \cite{lm, mathur} we postulate that the size
of the stretched horizon is determined by the same length scale that
sets the size of a dual-giant. Therefore we propose to place the
stretched horizon at:
\begin{equation}
\label{shsize}
r_0 = \frac{c_0 L}{\sqrt{N}}
\end{equation}
where $c_0$ is a pure numerical coefficient. 

Substituting $q_1= Q$ $\mu = q_2 = q_3 =0$ into (\ref{fivedbhone}) 
the 5-dimensional metric of the single charge superstar reads
\begin{equation}
\label{fivedmetric}
ds^2_5 = - H^{-2/3} \left( 1+\frac{r^2}{L^2} H \right) dt^2 + H^{1/3}
\left[ \frac{dr^2}{1+\frac{r^2}{L^2} H} + r^2 d\Omega_3^2 \right] 
\end{equation}
where $H = 1 + \frac{Q}{r^2}$. We assume that $Q_0 = Q/L^2 << 1$ so
that the metric is asymptotically globally $AdS_5$. The metric on a
space-like surface at the position $r = r_0$ is:
\begin{equation}
ds^3_{horizon} = \left( 1 + \frac{L^2 Q_0}{r_0^2} \right)^{1/3} r_0^2
d\Omega_3^2 
\end{equation}
where $d\Omega_3^2$ is the metric on a unit 3-sphere $S^3$. The area
of the stretched horizon becomes:
\begin{equation}
A \approx c_0^2 \, {\rm Vol}_{S^3} ~ \frac{L^3 Q_0^{1/2}}{N}.
\end{equation}
The 5-dimensional Newton's constant is given by:
\begin{equation}
G_N = \frac{8 \pi^3 g_s^2 \alpha'^4}{L^5} \sim \frac{L^3}{N^2}
\end{equation}
Then using the Bekenstein-Hawking entropy formula we get:
\begin{equation}
S_{BH} \sim \frac{A_{sh}}{G_N} \sim c_0^2 \, 
\left( \frac{L^4}{N g_s \alpha'^2} \right)^2 ~ N \sqrt{Q_0} \sim {\cal
C}_0 \, N \sqrt{Q_0}. 
\end{equation}
Up to a numerical coefficient this precisely matches with the entropy
prediction (\ref{theprediction}) of the gauge and string theory
microstate counting.

\section{M-theory examples}

Let us also consider similar singular geometries in the M-theory
examples $AdS_4\times S^7$ and $AdS_7\times S^4$. An analysis on lines
of \cite{mt} was carried out in \cite{bn,lmp}.

\subsection{Superstar in $AdS_4 \times S^7$}

The black hole in this case can carry four charges. We again restrict
ourselves to the single charge extremal limit. The 4 dimensional
solution is \cite{dl, sabra}:
\begin{eqnarray}
ds^2_4 &=& - H^{1/2} f \, dt^2 + H^{1/2} (f^{-1} dr^2 + r^2
d\Omega_2^2), \cr
X_1 &=& H^{-3/4}, ~~ X_2 = X_3 = X_4 = H^{1/4}, ~~ A = (1- H^{-1}) \,
dt 
\end{eqnarray}
where 
\begin{equation}
f = 1 + \frac{r^2}{L_4^2}H, ~~ H = 1 + \frac{Q}{r}.
\end{equation}
This solution also has a null singularity at $r = 0$ and admits an
interpretation \cite{bn,lmp} as a backreaction of a source of giant
gravitons now made out of $M5$ branes wrapping some $S^5 \subset S^7$
and carrying angular momentum. We refer the reader to \cite{bn,lmp}
for the details of this analysis and be content here with a
summary. The total number of giants is
\begin{equation}
n = 2^{1/2} N^{1/2} \frac{Q}{L_4}
\end{equation}
and the total angular momentum of the solution is:
\begin{equation}
\label{mthpone}
P_{\xi_1} = \frac{2 N^{3/2}}{3 \sqrt{2}} \frac{Q}{L_4}.
\end{equation}
The 4-dimensional Newton's constant is:
\begin{equation}
G_4 \sim \frac{L_4^2}{N^{3/2}}.
\end{equation}
There exist dual-giants made of $M2$ branes wrapping the $S^2 \subset
AdS_4$ and carrying angular momentum in $S^7$. The sizes of these are
\cite{gmt}: 
\begin{equation}
r = \frac{L_4}{\sqrt{N}} \frac{P_{\xi_1}}{2}.
\end{equation}
Using the hypothesis that the stretched horizon is to be placed at a
length scale decided by the size of the dual-giant we set:
\begin{equation}
r_0 = \frac{c_0 \, L_4}{\sqrt{N}}.
\end{equation}
where $c_0$ is again a pure number. Using the metric above it is easy
to calculate the area of this horizon and we find:
\begin{equation}
A_{sh} \sim \frac{c_0 L_4^2}{N^{3/4}} \sqrt{Q_0}
\end{equation}
where we have defined $Q_0 = Q/L_4$. This implies:
\begin{equation}
\label{mthentropyone}
S^{(4)}_{bh} \sim \frac{A_{sh}}{G_4} \sim N^{3/4} \sqrt{Q_0}.
\end{equation}
Even though there is no Lagrangian description for the dual field
theory to get a prediction for this entropy one can treat the chiral
primaries again using the picture of fermions in a harmonic oscillator
potential. One can also do a counting of giants or dual-giants on the
lines of Section 4.2 and the answer for the partition function comes
out similar to Eq.\,(\ref{sprediction}).  Therefore the prediction
from the CFT and the M-theory side for the entropy of our black hole
is again:
\begin{equation}
\label{fiveseven}
S = {\cal C} \sqrt{\Delta} + ...
\end{equation}
where $\cdots$ again mean corrections which are small in large-$N$ and
large-$\Delta$ limit. Substituting $P_{\xi_1}$ from
Eq.\,(\ref{mthpone}) for $\Delta$ in Eq.\,(\ref{fiveseven}) gives
precisely the same functional dependence of the entropy
(\ref{mthentropyone}) on the charge $Q_0$ and $N$.

\subsection{Superstar in $AdS_7 \times S^4$}

There exists a similar null singularity with 16 supercharges in this
background as well. The solution in 7 dimensions is \cite{dl, sabra}:
\begin{eqnarray}
ds^2_7 &=& - H^{-4/5} f \, dt^2 + H^{1/5} (f^{-1} dr^2 + r^2
d\Omega_5^2), \cr
X_1 &=& H^{-3/5} , ~~ X_2 = H^{2/5}, ~~ A = (1 - H^{-1}) \, dt
\end{eqnarray}
where 
\begin{equation}
f = 1 + \frac{r^2}{L_7^2}H, ~~ H = 1 + \frac{Q}{r^4}.
\end{equation}
The total charge of this solution is $P_{\xi_1} = (2N^3 Q)/(3L_7^4)
\sim N^3 Q_0$ where $Q_0 = Q/L_7^4$. The size of a dual-giant is $r
\sim P_{\xi_1}^{1/4} (L_7/\sqrt{N})$ and the 7-dimensional Newton's
constant is $G_7 \sim L_7^5/N^3$. Using the same hypothesis that the
position of the stretched horizon is determined by the length scale
that determines the size of the dual-giant $L_7/\sqrt{N}$, we again
take $r_0 = c_0 \frac{L_7}{\sqrt{N}}$ to be the horizon radius. Then
the entropy works out to be:
\begin{equation}
S^{(7)}_{bh} \sim N^{3/2} \sqrt{Q_0}
\end{equation}
which again matches (up to non-vanishing numerical factors) with the
prediction coming from counting the corresponding giant configurations
with the total angular momentum $P_{\xi_1} =(2N^3 Q)/(3L_7^4) \sim N^3
Q_0$. Thus we conclude that the null singularities of M-theory should
also be genuine black holes at the quantum level.
%\newpage

\section{More consequences}

Let us state some of the consequences of the duality considered in
Section 3. Suppose we start with the configuration of all $N$ fermions
being in their ground state. This represents the vacuum state of the
dual geometry, namely, empty $AdS_5 \times S^5$
\cite{berenstein,llm}. Now consider a generic fermion excitation. This
is represented by AdS/CFT {\it either} by a configuration of
dual-giants {\it or} by a configuration of giants. But both correspond
to describing the same fermion system either in terms of particles or
in terms of holes both of which should be equivalent. This leads to
the statement that a given giant configuration is equivalent to a
corresponding dual-giant configuration. However similar to the fermion
system the corresponding supergravity solutions would also have to be
identical.

Another consequence is the following. Again start with a giant
graviton configuration labeled by ${\vec b}_1 = (r_1, r_2, \cdots,
r_N)$. Now consider adding a dual-giant to this system. Since giants
are holes in the fermion picture the topmost fermion is away from its
ground state position $N-1$ and this distance is given by the number
of holes. As a consequence to excite the k$^{th}$ fermion we have to
have
\begin{equation}
s_k >
\sum_{i=k}^N r_i.
\end{equation}
For $k=1$ this implies that a dual-giant will have a non-zero radius
only when its angular momentum $s_1$ exceeds the total number
$\sum_{i=1}^N r_i$ of the giant gravitons. Let us test this prediction
in a simple context. 

To test this prediction we generically need the backreacted geometry
of the giant configuration under consideration. Below we work with the
single charge superstar which contains a total of $N Q_0$
D3-branes. So if we try to place a dual-giant D3-brane as a probe in
this background it should not have a non-zero radius unless the
angular momentum of the dual-giant exceeds $N Q_0$.

Fortunately the probe brane analysis in question has already been
studied in \cite{mt}. We summarise the results here. For a given
angular momentum $P_{\xi_i}$ in $S^5$ the equations of motion for the
world-volume theory are satisfied if the radius $r$ and the angles
$\alpha$, $\beta$ satisfy the following relations:
\begin{eqnarray}
\label{dthreeprobe}
\frac{r^2}{L^2} &=& \sum_{i=1}^3 \left( \frac{P_{\xi_i}}{N} -
\frac{q_i \mu_i^2}{L^2} \right), \cr
\mu_i^2 &=& \frac{\frac{P_{\xi_i}}{N} - \frac{q_i
\mu_i^2}{L^2}}{\sum_{j=1}^3 \left( \frac{P_{\xi_j}}{N} - 
\frac{q_j \mu_j^2}{L^2} \right) }.
\end{eqnarray}
Let us specialise to $P_{\xi_2} = P_{\xi_3} = 0$ and $q_2 = q_3 =
0$. Then the second of the equations (\ref{dthreeprobe}) can be solved
if $\mu_1 = 1$ and $\mu_2 = \mu_3 =0$. That is $\alpha =
\pi/2$. Substituting these into the first equation gives:
\begin{equation}
\frac{r^2}{L^2} = \frac{P_{\xi_1}}{N} -
\frac{q_1}{L^2} 
\end{equation}
That is, the typical radius at which the dual-giant settles down is:
\begin{equation}
r = \frac{L}{\sqrt{N}} \, \left[ P_{\xi_1} -
\frac{N q_1}{L^2} \right]^{1/2} = \frac{L}{\sqrt{N}} \, \left[
  P_{\xi_1} - N Q_0 \right]
\end{equation}
Thus we see that the dual-giant will have non-vanishing radius only
when its angular momentum exceeds the total number of giants $N Q_0$
as predicted. Similar predictions apply for the M-theory cases as well
and one can verify them using the probe brane analyses of
\cite{lmp}. One should be able to verify this prediction further by
probing various supergravity solutions presented in \cite{llm} with
D3-brane configurations with angular momenta. We will not do this here
however.

It follows from Section 3 that $M$ giant gravitons with angular
momentum $N$ are dual to $N$ dual-giants with $M$ units of angular
momentum each. However depending on how big the values of $M$ and $N$
are relative to each other it should be possible to think of either
$M$ or $N$ as the number of giants/dual-giants in the supergravity
background of the other. A similar `correspondence principle' was
proposed in \cite{bn}.

\section{Conclusion}

In this paper we reconsidered the 1/2-BPS configurations on both the
gauge theory and string theory sides of the AdS/CFT correspondence. We
have argued that there is an upper limit on the number of dual-giants
one can place in $AdS_5 \times S^5$. Using the fermion picture that
arises from the matrix model truncation of the gauge theory we have
set up a mapping between the configurations of fermions, giants and
the dual-giants. This leads to a duality map between giant graviton
configurations and the dual-giant graviton configurations which is
hinted at by \cite{msv} in the M-theory context. Some consequences of
the mapping proposed have also been tested in simple cases.

Further, we found the partition function of the 1/2-BPS states both by
counting the chiral primaries using the matrix model and various giant
graviton configurations which match. The density of 1/2-BPS states is
shown to grow exponentially with the total R-charge. Then we proposed
that this density of states should be carried by the single charge
`superstar' geometry of \cite{mt}. A stretched horizon was proposed
and shown to reproduce the entropy predicted by both the CFT and the
string theory microstate counting.

It will be interesting if one can argue more concretely for the
conjecture that the superstar carries the entropy coming from the
degeneracy of chiral primaries on the lines of
\cite{dabholkar,dkm}. This will require Wald's formula \cite{wald} for
entropy applied to gauged supergravities.

In recent times a new understanding of the notion of stretched horizon
has been provided by Mathur {\it et al} (see for instance \cite{lm}).
It will be interesting to see if a similar picture to the D1-D5 system
emerges even for the geometries considered here. In particular, one
should be able to extract the relevant microstate geometries from
those found recently in \cite{llm}. One must then be able to
`coarse-grain' over these geometries and get the entropy formula. See
\cite{buchel} for a related discussion.

A class of supersymmetric black holes in $AdS_5$ with finite area
horizons of spherical topology have been discovered recently
\cite{grone,grtwo}. When lifted to 10 dimensions they preserve just
two supersymmetries \cite{ggs}. Of course understanding their
entropies from the CFT side remains an important outstanding
problem. See \cite{gh} for a discussion on the microstate counting of
the near extremal R-charge black holes in AdS.

\section*{Acknowledgments}

It is my pleasure to thank Jaume Gomis, Gautam Mandal, Rob Myers and
Adam Ritz for useful discussions and helpful comments on the draft. I
am also thankful to Alex Buchel for a discussion on \cite{buchel} and
Andrei Starinets for interest and collaboration.


\begin{thebibliography}{99}

%\cite{Maldacena:1997re}
%\bibitem{Maldacena:1997re}
\bibitem{maldacena}
J.~M.~Maldacena,
``The large N limit of superconformal field theories and supergravity,''
Adv.\ Theor.\ Math.\ Phys.\  {\bf 2}, 231 (1998)
[Int.\ J.\ Theor.\ Phys.\  {\bf 38}, 1113 (1999)]
[arXiv:hep-th/9711200].
%%CITATION = HEP-TH 9711200;%%

%\cite{Aharony:1999ti}
%\bibitem{Aharony:1999ti}
\bibitem{agmoo}
O.~Aharony, S.~S.~Gubser, J.~M.~Maldacena, H.~Ooguri and Y.~Oz,
``Large N field theories, string theory and gravity,''
Phys.\ Rept.\  {\bf 323}, 183 (2000)
[arXiv:hep-th/9905111].
%%CITATION = HEP-TH 9905111;%%

%\cite{McGreevy:2000cw}
%\bibitem{McGreevy:2000cw}
\bibitem{mst}
J.~McGreevy, L.~Susskind and N.~Toumbas,
``Invasion of the giant gravitons from anti-de Sitter space,''
JHEP {\bf 0006}, 008 (2000)
[arXiv:hep-th/0003075].
%%CITATION = HEP-TH 0003075;%%

%\cite{Grisaru:2000zn}
%\bibitem{Grisaru:2000zn}
\bibitem{gmt}
M.~T.~Grisaru, R.~C.~Myers and O.~Tafjord,
``SUSY and Goliath,''
JHEP {\bf 0008}, 040 (2000)
[arXiv:hep-th/0008015].
%%CITATION = HEP-TH 0008015;%%

%\cite{Hashimoto:2000zp}
%\bibitem{Hashimoto:2000zp}
\bibitem{hhi}
A.~Hashimoto, S.~Hirano and N.~Itzhaki,
``Large branes in AdS and their field theory dual,''
JHEP {\bf 0008}, 051 (2000)
[arXiv:hep-th/0008016].
%%CITATION = HEP-TH 0008016;%%

%\cite{Lin:2004nb}
%\bibitem{Lin:2004nb}
\bibitem{llm}
H.~Lin, O.~Lunin and J.~Maldacena,
``Bubbling AdS space and 1/2 BPS geometries,''
arXiv:hep-th/0409174.
%%CITATION = HEP-TH 0409174;%%

%\cite{Corley:2001zk}
%\bibitem{Corley:2001zk}
\bibitem{cjr}
S.~Corley, A.~Jevicki and S.~Ramgoolam,
``Exact correlators of giant gravitons from dual N = 4 SYM theory,'' 
Adv.\ Theor.\ Math.\ Phys.\  {\bf 5}, 809 (2002)
[arXiv:hep-th/0111222].
%%CITATION = HEP-TH 0111222;%%

%\cite{Berenstein:2004kk}
%\bibitem{Berenstein:2004kk}
\bibitem{berenstein}
D.~Berenstein,
``A toy model for the AdS/CFT correspondence,''
JHEP {\bf 0407}, 018 (2004)
[arXiv:hep-th/0403110].
%%CITATION = HEP-TH 0403110;%%

%\cite{Myers:2001aq}
%\bibitem{Myers:2001aq}
\bibitem{mt}
R.~C.~Myers and O.~Tafjord,
``Superstars and giant gravitons,''
JHEP {\bf 0111}, 009 (2001)
[arXiv:hep-th/0109127].
%%CITATION = HEP-TH 0109127;%%

%\cite{Behrndt:1998ns}
%\bibitem{Behrndt:1998ns}
\bibitem{bcs1}
K.~Behrndt, A.~H.~Chamseddine and W.~A.~Sabra,
``BPS black holes in N = 2 five dimensional AdS supergravity,''
Phys.\ Lett.\ B {\bf 442}, 97 (1998)
[arXiv:hep-th/9807187].
%%CITATION = HEP-TH 9807187;%%

%\cite{Behrndt:1998jd}
%\bibitem{Behrndt:1998jd}
\bibitem{bcs2}
K.~Behrndt, M.~Cvetic and W.~A.~Sabra,
``Non-extreme black holes of five dimensional N = 2 AdS
supergravity,'' 
Nucl.\ Phys.\ B {\bf 553}, 317 (1999)
[arXiv:hep-th/9810227].
%%CITATION = HEP-TH 9810227;%%

%\cite{Sen:1995in}
%\bibitem{Sen:1995in}
\bibitem{sen}
A.~Sen,
``Extremal black holes and elementary string states,''
Mod.\ Phys.\ Lett.\ A {\bf 10}, 2081 (1995)
[arXiv:hep-th/9504147].
%%CITATION = HEP-TH 9504147;%%

%\cite{Wald:1993nt}
%\bibitem{Wald:1993nt}
\bibitem{wald}
R.~M.~Wald,
``Black hole entropy in the Noether charge,''
Phys.\ Rev.\ D {\bf 48}, 3427 (1993)
[arXiv:gr-qc/9307038].
%%CITATION = GR-QC 9307038;%%

%\cite{LopesCardoso:1998wt}
%\bibitem{LopesCardoso:1998wt}
\bibitem{cdm1}
G.~Lopes Cardoso, B.~de Wit and T.~Mohaupt,
``Corrections to macroscopic supersymmetric black-hole entropy,''
Phys.\ Lett.\ B {\bf 451}, 309 (1999)
[arXiv:hep-th/9812082].
%%CITATION = HEP-TH 9812082;%%

%\cite{LopesCardoso:1999cv}
%\bibitem{LopesCardoso:1999cv}
\bibitem{cdm2}
G.~Lopes Cardoso, B.~de Wit and T.~Mohaupt,
``Deviations from the area law for supersymmetric black holes,''
Fortsch.\ Phys.\  {\bf 48}, 49 (2000)
[arXiv:hep-th/9904005].
%%CITATION = HEP-TH 9904005;%%

%\cite{Dabholkar:2004yr}
%\bibitem{Dabholkar:2004yr}
\bibitem{dabholkar}
A.~Dabholkar,
``Exact counting of black hole microstates,''
arXiv:hep-th/0409148.
%%CITATION = HEP-TH 0409148;%%

%\cite{Dabholkar:2004dq}
%\bibitem{Dabholkar:2004dq}
\bibitem{dkm}
A.~Dabholkar, R.~Kallosh and A.~Maloney,
``A Stringy Cloak for a Classical Singularity,''
arXiv:hep-th/0410076.
%%CITATION = HEP-TH 0410076;%%

%\cite{Lunin:2002qf}
%\bibitem{Lunin:2002qf}
\bibitem{lm}
O.~Lunin and S.~D.~Mathur,
``Statistical interpretation of Bekenstein entropy for systems with a 
stretched horizon,''
Phys.\ Rev.\ Lett.\  {\bf 88}, 211303 (2002)
[arXiv:hep-th/0202072].
%%CITATION = HEP-TH 0202072;%%

%\cite{Mathur:2004sv}
%\bibitem{Mathur:2004sv}
\bibitem{mathur}
S.~D.~Mathur,
``Where are the states of a black hole?,''
arXiv:hep-th/0401115.
%%CITATION = HEP-TH 0401115;%%

%\cite{Balasubramanian:2001dx}
%\bibitem{Balasubramanian:2001dx}
\bibitem{bn}
V.~Balasubramanian and A.~Naqvi,
``Giant gravitons and a correspondence principle,''
Phys.\ Lett.\ B {\bf 528}, 111 (2002)
[arXiv:hep-th/0111163].
%%CITATION = HEP-TH 0111163;%%

%\cite{Leblond:2001gn}
%\bibitem{Leblond:2001gn}
\bibitem{lmp}
F.~Leblond, R.~C.~Myers and D.~C.~Page,
``Superstars and giant gravitons in M-theory,''
JHEP {\bf 0201}, 026 (2002)
[arXiv:hep-th/0111178].
%%CITATION = HEP-TH 0111178;%%

%\cite{Bena:2004qv}
%\bibitem{Bena:2004qv}
\bibitem{bs}
I.~Bena and D.~J.~Smith,
``Towards the solution to the giant graviton puzzle,''
arXiv:hep-th/0401173.
%%CITATION = HEP-TH 0401173;%%

%\cite{deMelloKoch:2004ws}
%\bibitem{deMelloKoch:2004ws}
\bibitem{kg}
R.~de Mello Koch and R.~Gwyn,
``Giant graviton correlators from dual SU(N) super Yang-Mills theory,''
JHEP {\bf 0411}, 081 (2004)
[arXiv:hep-th/0410236].
%%CITATION = HEP-TH 0410236;%%

%\cite{Caldarelli:2004ig}
%\bibitem{Caldarelli:2004ig}
\bibitem{cs}
M.~M.~Caldarelli and P.~J.~Silva,
``Giant gravitons in AdS/CFT. I: Matrix model and back reaction,''
JHEP {\bf 0408}, 029 (2004)
[arXiv:hep-th/0406096].
%%CITATION = HEP-TH 0406096;%%

%\cite{Itzhaki:2004te}
%\bibitem{Itzhaki:2004te}
\bibitem{im}
N.~Itzhaki and J.~McGreevy,
``The large N harmonic oscillator as a string theory,''
arXiv:hep-th/0408180.
%%CITATION = HEP-TH 0408180;%%

%\cite{Mikhailov:2000ya}
%\bibitem{Mikhailov:2000ya}
\bibitem{mikhailov}
A.~Mikhailov,
``Giant gravitons from holomorphic surfaces,''
JHEP {\bf 0011}, 027 (2000)
[arXiv:hep-th/0010206].
%%CITATION = HEP-TH 0010206;%%

%\cite{Balasubramanian:2001nh}
%\bibitem{Balasubramanian:2001nh}
\bibitem{bbns}
V.~Balasubramanian, M.~Berkooz, A.~Naqvi and M.~J.~Strassler,
``Giant gravitons in conformal field theory,''
JHEP {\bf 0204}, 034 (2002)
[arXiv:hep-th/0107119].
%%CITATION = HEP-TH 0107119;%%

%\cite{Maldacena:2002rb}
%\bibitem{Maldacena:2002rb}
\bibitem{msv}
J.~Maldacena, M.~M.~Sheikh-Jabbari and M.~Van Raamsdonk,
``Transverse fivebranes in matrix theory,''
JHEP {\bf 0301}, 038 (2003)
[arXiv:hep-th/0211139].
%%CITATION = HEP-TH 0211139;%%

%\cite{Das:2000st}
%\bibitem{Das:2000st}
\bibitem{djm}
S.~R.~Das, A.~Jevicki and S.~D.~Mathur,
``Vibration modes of giant gravitons,''
Phys.\ Rev.\ D {\bf 63}, 024013 (2001)
[arXiv:hep-th/0009019].
%%CITATION = HEP-TH 0009019;%%

%\cite{Cvetic:1999xp}
%\bibitem{Cvetic:1999xp}
\bibitem{cdhjl3mpst}
M.~Cvetic {\it et al.},
``Embedding AdS black holes in ten and eleven dimensions,''
Nucl.\ Phys.\ B {\bf 558}, 96 (1999)
[arXiv:hep-th/9903214].
%%CITATION = HEP-TH 9903214;%%

%\cite{Duff:1999gh}
%\bibitem{Duff:1999gh}
\bibitem{dl}
M.~J.~Duff and J.~T.~Liu,
``Anti-de Sitter black holes in gauged N = 8 supergravity,''
Nucl.\ Phys.\ B {\bf 554}, 237 (1999)
[arXiv:hep-th/9901149].
%%CITATION = HEP-TH 9901149;%%

%\cite{Sabra:1999ux}
%\bibitem{Sabra:1999ux}
\bibitem{sabra}
W.~A.~Sabra,
``Anti-de Sitter BPS black holes in N = 2 gauged supergravity,''
Phys.\ Lett.\ B {\bf 458}, 36 (1999)
[arXiv:hep-th/9903143].
%%CITATION = HEP-TH 9903143;%%

%\cite{Buchel:2004mc}
%\bibitem{Buchel:2004mc}
\bibitem{buchel}
A.~Buchel,
``Coarse-graining 1/2 BPS geometries of type IIB supergravity,''
arXiv:hep-th/0409271.
%%CITATION = HEP-TH 0409271;%%

%\cite{Gutowski:2004ez}
%\bibitem{Gutowski:2004ez}
\bibitem{grone}
J.~B.~Gutowski and H.~S.~Reall,
``Supersymmetric AdS(5) black holes,''
JHEP {\bf 0402}, 006 (2004)
[arXiv:hep-th/0401042].
%%CITATION = HEP-TH 0401042;%%

%\cite{Gutowski:2004yv}
%\bibitem{Gutowski:2004yv}
\bibitem{grtwo}
J.~B.~Gutowski and H.~S.~Reall,
``General supersymmetric AdS(5) black holes,''
JHEP {\bf 0404}, 048 (2004)
[arXiv:hep-th/0401129].
%%CITATION = HEP-TH 0401129;%%

%\cite{Gauntlett:2004cm}
%\bibitem{Gauntlett:2004cm}
\bibitem{ggs}
J.~P.~Gauntlett, J.~B.~Gutowski and N.~V.~Suryanarayana,
``A deformation of AdS$_5 \times$ S$^5$,''
Class.\ Quant.\ Grav.\  {\bf 21}, 5021 (2004)
[arXiv:hep-th/0406188].
%%CITATION = HEP-TH 0406188;%%

%\cite{Gubser:2004xx}
%\bibitem{Gubser:2004xx}
\bibitem{gh}
S.~S.~Gubser and J.~J.~Heckman,
``Thermodynamics of R-charged black holes in AdS(5) from effective strings,''
arXiv:hep-th/0411001.
%%CITATION = HEP-TH 0411001;%%



\end{thebibliography}
\end{document}